\documentstyle[preprint,aps]{revtex}

\begin{document}

\widetext
\draft
\title {Elastic turbulence in a polymer solution flow}
\author {Alexander Groisman and Victor Steinberg}
\address {Department of Physics of Complex Systems, Weizmann Institute of Science, Rehovot 76100, Israel}
\date{\today}
\maketitle


{\bf Turbulence is one of the most fascinating phenomena in nature and one of the
biggest challenges for modern physics. It is common knowledge that a flow of
a simple, Newtonian fluid is likely to be turbulent, when velocity is high, viscosity
is low and size of the tank is large\cite{Landau,Tritt}. Solutions of flexible long-chain
polymers are known as visco-elastic fluids\cite{bird}. In our experiments we show, that
flow of a polymer solution with large enough elasticity can become quite
irregular even at low velocity, high viscosity and in a small tank. The
fluid motion is excited in a broad range of spatial and temporal scales. The
flow resistance increases by a factor of about twenty. So, while the Reynolds
number, $\boldmath{Re}$, may be arbitrary low, the observed flow has all main features of
developed turbulence, and can be compared to turbulent flow in a pipe at
$\bf {Re\simeq 10^5}$\cite{Landau,Tritt}. This {\it elastic turbulence} is accompanied by
significant stretching of the polymer molecules, and the resulting increase
of the elastic stresses can reach two orders of magnitude.}

Motion of simple, low molecular, {\it Newtonian} fluids is governed by
the Navier-Stokes equation \cite{Landau,Tritt}. This equation has a non-linear term, which
is inertial in its nature. The ratio between the non-linearity and viscous
dissipation is given by the Reynolds number, $Re=VL/\nu $,where $V$ is
velocity, $L$ is characteristic size and $\nu$ is kinematic viscosity of the fluid.
When $Re$ is high, non-linear effects are strong and the flow is likely to be turbulent.
So, turbulence is a paradigm for a strongly non-linear phenomenon\cite{Landau,Tritt}.

Solutions of flexible high molecular weight polymers differ from Newtonian fluids
in many aspects \cite{bird}. The most striking elastic property of the polymer
solutions is that stress does not immediately become zero
when the fluid motion stops, but rather decays with some
characteristic time, $\lambda $, which can reach seconds and even minutes.
Equation of motion for dilute polymer solutions differs from the Navier-Stokes
equation by an additional linear term due to the elastic stress, $\bf {\tau }$ \cite{bird}.
Since the elastic stress is caused by stretching of the
polymer coils, it depends on history of motion
and deformation of fluid elements along its flow trajectory. This implies
non-linear relationship between $\bf {\tau }$ and the rate of deformation
in a flow \cite{bird}. The {\it non-linear mechanical properties} of polymer solutions
are well manifested in their large extensional viscosity at high rates of extension
\cite{extension} and in the Weissenberg effect \cite{Weiss,bird}. Degree of non-linearity
in the mechanical properties is expressed by the Weissenberg number,
$Wi=V\lambda /L$, which is a product of
characteristic rate of deformation and the relaxation time, $\lambda$.

It is reasonable to inquire, whether non-linearity of mechanical properties
of a fluid can give rise to turbulent flow, when the equation of motion is linear.
For a polymer solution this corresponds to a state, when the Weissenberg number is
large, while the Reynolds number is small. This situation
can be realized, if the parameter of elasticity $Wi/Re=\lambda\nu/L^{2}$ is large
enough. An important step in investigation of influence of the non-linear mechanical
properties on flow was made about a decade ago, when {\it purely elastic}
instability was experimentally identified in curvilinear shear flows \cite{magda,LMS}.
This instability occurs at moderate $Wi$ and vanishingly small $Re$ and
is driven by the elastic stresses \cite{LMS,Ours}.
As a result of the instability,
secondary, in general oscillatory, vortex flows develop, and flow resistance
somewhat increases \cite{magda,LMS,LMS2,Ours,McK}. Flow instabilities in elastic liquids
are reviewed in \cite{Larson,Shaq}.

There is no unique commonly accepted definition of turbulent flow \cite{Tritt}, so it
is usually identified by its major features \cite{Landau,Tritt}.
Turbulence implies fluid motion in a broad range of
spatial and temporal scales, so that many degrees of freedom are excited in
the system. A striking practically important characteristic of turbulent flows is
major increase in the flow resistance compared to an imaginary laminar flow
with the same $Re$. Experiments that we report in this
letter show, how these main features of turbulence appear in a flow of a
highly elastic polymer solution at low Reynolds numbers.

For our experiments we chose swirling flow between two parallel disks,
Fig.1, and a dilute solution of high molecular weight polyacrylamide in a
viscous sugar syrup, as the working fluid. The curvature ratio was made quite
high, $d/R=0.263$, in order to provide destabilization of the primary shear flow
and development of the secondary vortical fluid motion at lower shear rates \cite{LMS,McK}.
(The flow between two plates with small $d/R$ was studied before in context of the
purely elastic instability \cite{McK}.)
The whole flow set-up was mounted on top of a commercial viscometer (AR-1000 of TA-instruments),
so that
we could precisely measure the angular velocity, $\omega $, of the rotating
upper plate and the torque applied to it. In this way we were able to estimate the average
shear stress, $\sigma$, in the polymer solution and to compare it with the stress
in the laminar flow, $\sigma_{lam}$, with the same applied shear rate.
In Newtonian fluids the  ratio $\sigma/\sigma_{lam}$ generally grows with $Re$ as the flow
becomes increasingly irregular, and the magnituide of $\sigma/\sigma_{lam}$ can be considered
as a measure of strength of turbulence and turbulent resistance. In our
set-up $\sigma$ becomes 30\% higher than $\sigma_{lam}$ at $Re=70$, that can be
regarded as a point when inertial effects become significant.

Dependence of $\sigma/\sigma_{lam}$ on the shear rate, $\dot{\gamma}=\omega
R/d $, for flow of the polymer solution in the experimental system is shown
in Fig.2 (first curve). One can see that at $\dot{\gamma}$ of about 1 s$^{-1}$
(corresponding to $Wi\equiv \lambda\dot{\gamma}=3.5$) a sharp transition occurs that
appears as a significant
increase in the apparent viscosity. The Reynolds number at the transition
point is about 0.3, so that the inertial effects are quite negligible there. The
transition has pronounced hysteresis, which is typical for the purely elastic
flow instability \cite{Ours}. The ratio $\sigma/\sigma_{lam}$ keeps growing with the shear rate
and at the highest $\dot{\gamma}$, that has been reached, the flow resistance is
about 12 times larger than in the laminar flow.
In the same range of shear rates, flow of the pure solvent is completely
laminar and the ratio $\sigma/\sigma_{lam}$ is unity within resolution of the viscometer (about
1\%). To make sure, that the observed flow phenomena were indeed caused by
the solution elasticity, we measured $\sigma$ for a few solutions with
the same polymer concentration, but different relaxation times, $\lambda $.
The curves of $\sigma/\sigma_{lam}$ coincided, when plotted against $Wi$,
while the Reynolds number turned out to be completely irrelevant (see also \cite{Ours}).
The growth of the resistance in the polymer solution flow becomes even larger, when the
size of the gap is increased (2nd curve in Fig.2). Then the ratio $\sigma/\sigma_{lam}$
reaches a value of 19. Such growth of the flow resistance
is found for Newtonian fluids in the same flow geometry at $Re$ of about 2$%
\cdot $10$^{4}$. For flow in a circular pipe this value of $\sigma/\sigma_{lam}$
is reached at $Re\simeq 10^{5}$, which is usually considered
as a region of rather developed turbulence \cite{Landau}.

Two representative snapshots of the polymer solution flow above the transition (at $\dot{\gamma}=4$ s$%
^{-1}$) are shown in Fig.3. One can see that the flow patterns are very much
irregular and structures of quite different sizes appear. This visual
impression is confirmed by a more careful analysis. In Fig. 4 one can see
average Fourier spectra of the brightness profiles along the diameter and
along the circumference. The both spectra exhibit power law decay over a decade
in the vawenumber domain.

Characteristic time spectra of velocity fluctuations at different shear
rates are shown in Fig. 5. Flow velocity was measured in the horizontal
plane in the center of the set-up, where its average value is zero. As the
shear rate is raised, the power of fluctuations increases and characteristic
frequencies become higher. On the other hand, the general form of the spectra
remains very much the same. In particular, just like for the spatial
spectra in Fig. 4, there is a region of a power law decay, which spans
over about a decade in frequencies. This power law dependence in the broad
ranges of spatial and temporal frequencies actually means that the fluid
motion is excited at all those spatial and temporal scales. Spectra of
radial and azimuthal velocities taken at different points with non-zero
average flow had the same general appearance and close values of exponents
in the power law decay range.

Summarizing all the results, we conclude that the flow of the elastic
polymer solution at sufficiently high $Wi$ has indeed all main features of
developed turbulence. By the strength of the turbulent resistance, and by
the span of scales in space and time, where the fluid motion is
excited, the observed flow can be compared to turbulence in a Newtonian fluid
in a pipe at $Re$ of about 10$^{5}$. This apparently
turbulent flow arises solely due to non-linear mechanical properties
of the elastic polymer solutions. So, we called the phenomenon
{\it elastic turbulence}, in contrast to the usual
inertial turbulence which is observed in Newtonian fluids at high $%
Re$. (The name "elastic turbulence" has been used before for designation of apparently
disordered flows in polymeric liquids \cite{Gies}. No attempt has been ever made,however,
to characterize those flows quantitatively.)

The elastic turbulence has many features that are in sharp contradiction to
the intuition based on the Newtonian fluid mechanics. So, velocity required
for excitation of inertial turbulence in a Newtonian fluid is proportional to the fluid
viscosity. On the other hand, the polymer relaxation time, $\lambda $,
usually grows proportionally to the viscosity. Since at constant $d/R$ transition to
elastic turbulence occurs at a certain value of the Weissenberg number, $Wi=V\lambda /L$,
choosing more viscous polymer solutions, one can excite
turbulence at lower velocities. Indeed, using a solution of polymers in a
very viscous sugar syrup, we observed transition to the elastic turbulence
at a rotation rate of 0.05 s$^{-1}$ (corresponding to $Re=10^{-2}$).
Further, in an elastic polymer solution the scale of time, $\lambda $,
does not depend on the size of
the system. Therefore, as long as the ratio $d/R$
is preserved, transition to turbulence should occur at the same $\omega$ and
the dependence of $\sigma/\sigma_{lam}$ on $\dot{\gamma}$ should not change with
the size of the system. We repeated the measurements of $\sigma$
in a small set-up having all the dimensions reduced by a factor of 4
compared to the standard system. The dependence of $\sigma/\sigma_{lam}$ on
$\dot{\gamma}$ was found to be the same (the data are not shown) as in Fig.2,
while the characteristic velocities and the Reynolds numbers were
lower by factors of 4 and 16, respectively. Therefore, we
believe that using polymer solutions with sufficiently high elasticity, we can
excite turbulent motion at arbitrary low velocities and in arbitrary small
tanks. (The size of the tank still has to be large, compared to the
size of the polymer coils.)

An important question about the elastic turbulence is, where the turbulent
resistance comes from. In the inertial turbulence the origin of the large
resistance is the Reynolds stress,
which is connected with high kinetic energy of the turbulent motion and takes
major part in the momentum transfer in the flow. Elastic turbulence occurs
at low Reynolds numbers. From our velocity measurements in the standard
set-up, contribution of the Reynolds stress to the flow resistance could be
estimated as being less than 0.5\%. The contribution of the viscous shear stress
of the Newtonian solvent, averaged across the fluid layer, is always the same as in laminar
flow and cannot change. Thus, the whole increase in the flow resistance should be
due to the elastic stress. The data shown in Fig.2 (2nd curve) imply that the polymer
contribution to the stress increases by a factor of up to 65, compared to
laminar flow with the same average shear rate. This suggestion agrees very
well with our measurements of relaxation of the shear stress
after the fluid motion is stopped. The elastic part, $\tau$, of the whole stress, 
is identified by its slow relaxation with a characteristic time of the order $\lambda$.
In the elastic turbulence this slowly relaxing part can become two orders of magnitude
larger than in the laminar flow with the same shear rate.
This major growth of the elastic stress should be connected with
vast extension of the polymer molecules in the turbulent flow.

Thus, the scenario of development of the elastic turbulence is apparently the following.
The polymer molecules are stretched in the primary shear flow, drive it unstable
and cause irregular secondary flow.
The flow acts back on the polymer molecules stretching them further and becomes
increasingly turbulent, until a kind of saturated dynamic state is
finally reached. Density of the elastic energy of the stretched polymers
can be estimated as $Wi\cdot \tau /2$. So, it should increase in the elastic
turbulent flow by about the same factor as the elastic stress, $\tau $,
while the kinetic energy is always quite small.

{\bf Methods.}
We used solution of 65\% saccharose, 1\% NaCl in water, viscosity
$\eta_{s}=0.324$ Pa$\cdot$s, as a solvent for the polymer. We added
polyacrylamide (M$_{w}$=18,000,000, Polysciences) at a concentration of 80 ppm by weight.
The solution viscosity was $\eta=0.424$ Pa$\cdot$s at $\dot{\gamma}=1$ s$^{-1}$.
The relaxation time, $\lambda$, estimated from the phase shift between the stress and
the shear rate in oscillatory tests, was 3.4 s.

{\bf Figure captions}

Fig.1 Schematic drawing of the experimantal set-up. The set-up
consists of a stationary cylindrical cup with a plain bottom (the
lower plate), which is concentric with the rotating upper plate.
The latter is attached to the shaft of a commercial rheometer. The
radii of the upper and the lower plates are $R$=38 mm and
$R_{2}$=43.6 mm, respectively. The liquid is filled till a level
$d$, which is 10 mm, if not stated differently. The upper plate
just touches the surface of the liquid. A special cover is put
from above to minimize evaporation of the liquid. The temperature
is stabilized at 12 $^{\circ}$C by circulating water under the
steel lower plate. The walls of the cup are transparent that
allows to make laser Doppler velocimeter measurements by
collecting light scattared from the crossing point of two
horizontal laser beams. In the experimental runs, where the flow
has to be viewed from below, the lower plate is made from
plexiglass and a mirror tilted by 45$^{\circ}$ is placed under the
lower plate. Then the flow patterns are captured by a CCD camera
from a side and the temperature is stabilized by circulating air
in a closed box.

Fig.2 The ratio of the average stress, $\sigma$, measured in the
flow, to the stress, $\sigma_{lam}$, in the laminar flow with the
same boundary conditions, as a function of the shear rate,
$\dot{\gamma}$. The curves 1 and 2 are for the polymer solution
flow with the gap between the plates $d=10$ mm and $20$ mm,
respectively. The shear rate was gradually varied in time, very
slowly (by about 10\% per hour) in the transition region, and
faster below and above it. Thin black lines are for increasing
$\dot{\gamma}$, while thick gray lines correspond to lowering of
$\dot{\gamma}$. The curve 3 is for the pure solvent. Mechanical
degradation of the polymers was quite small at shear rates below
1.5 s$^{-1}$ and 1 s$^{-1}$ for $d=10$ mm and $20$ mm,
respectively. So, the dependences of $\sigma/\sigma_{lam}$ on
$\dot{\gamma}$ in those regions were reproducible in consecutive
runs within about 1\%. Degradation effects became appreciable at
higher shear rates, and elasticity typically decreased by up to
10\% as a result of the runs shown by the curves 1 and 2.

Fig.3 Two snapshots of the flow at $Wi=13$, $Re=0.7$. The flow
under the black upper plate is visualized by seeding the fluid
with light reflecting flakes (1\% of the Kalliroscope liquid). The
fluid is illuminated by ambient light. Although the pattern is
quite irregular, one can see that appearing structures tend to
have spiral-like form. The dark spot in the middle corresponds to
the center of a big persistent thoroidal vortex that has
dimensions of the whole set-up.

Fig.4 Average Fourier spectra of the brightness profiles taken
along the diameter (thin black line) and along the circumference
at a radius of $2d$ (thick gray line). The averaging was made over
long series of flow pattern snapshots taken in consecutive moments
of time. The wavelength is measured in units of $d$, so that the
wavenumber, $k$, of unity corresponds to a length of $2\pi d$. The
spectrum taken along the diameter apparently differs from the
azimuthal spectrum by a series of broad peaks. This may be a
manifestation of the fact, that the flow is not completely
structureless and homogeneous along the radial direction (see
Fig.3). The visualization method we used (Fig. 3) does not provide
direct information about the fluid velocity. So, the specific
value of the exponent in the power law fit, $A \sim k^{-1}$,
should not be given much consideration to.

Fig.5 Power spectra of velocity fluctuations in the standard
set-up at different shear rates, $\dot{\gamma}$. The fluid
velocity was measured by a laser Doppler velocimeter in the center
of the flow. The curves 1 - 5 correspond to $\dot{\gamma}$=1.25,
1.85, 2.7, 4, and 5.9 $s^{-1}$, respectively (all above the
transition point $\dot{\gamma}\simeq 1$, Fig.2). The power, P, of
fluctuations is fitted by a power law, $P \sim f^{-3.5}$, for
$\dot{\gamma}=4 s^{-1}$ over about a decade in frequencies, $f$.

 {\bf Acknowledgement.}
 Support of the Minerva Center for Nonlinear Physics of Complex Systems is
gratefully acknowledged.


\begin{references}

\bibitem{Landau} Landau L. D. and Lifschitz, E. M., {\it Fluid mechanics},
Pergamon Press , Oxford, 1987.

\bibitem{Tritt}Tritton D. J., {\it Physical Fluid Dynamics},
Clarendon Press, Oxford, 1988.

\bibitem{bird}  Bird, R. B., Curtiss, C. F., Armstrong, R. C., and
Hassager, O., {\it Dynamics of polymeric liquids}, John Wiley, NY, 1987.

\bibitem{extension} Tirtaatmadja, V. and Sridhar, T. A filament
stretching device for measurement of extensional viscosity,
{\it J. Rheology} {\bf 37}, 1081-1102 (1993);


\bibitem{Weiss} Weissenberg,  K., A continuum theory of rheological phenomena,
{\it Nature} {\bf 159}, 310-311 (1947).

\bibitem{magda} Magda, J. J. and Larson, R. G., A transition occuring in
ideal elastic liquids during shear flow, {\it J. Non-Newtonian Fluid Mech.}
{\bf 30}, 1-19 (1988).

\bibitem{LMS} Muller, S. J., Larson, R. G. and Shaqfeh, E. S. G. A purely
elastic transition in Taylor-Couette flow, {\it Rheol. Acta} {\bf 28},
499-503 (1989).

\bibitem{LMS2}Larson, R. G., Shaqfeh, E. S. G., and Muller, S.J.
A purely viscoelastic instability in Taylor-Couette flow, {\it J. Fluid Mech.} {\bf 218},
573-600 (1990).

\bibitem{Ours}Mechanism of elastic instability in Couette flow of polymer
solutions: experiment,{\it Phys. Fluids} {\bf 10}, 2451-2463 (1998).

\bibitem{McK} Byars J. A., \"{O}ztekin, A., Brown R. A. and McKinley G. H.,
Spiral instabilities in the flow of highly elastic fluids between rotating
parallel disks, {\it J. Fluid Mech.} {\bf 271}, 173-218 (1994)

\bibitem{Larson} Larson, R. G., Instabilities in viscoelastic flows,
{\it Rheol. Acta} {\bf 31}, 213-263, (1992)

\bibitem{Shaq} Shaqfeh, E. S. G., Purely elastic instabilities in viscometric flows,
{\it Annu. Rev. Fluid Mech.} {\bf 28}, 129-185 (1996).

\bibitem{Gies} Giesekus, H. W., Non-linear effects in the flow of visco-elastic fluids
through slits and holes, {\it Rheol. Acta} {\bf 7}, 127-138, (1968).



\end{references}
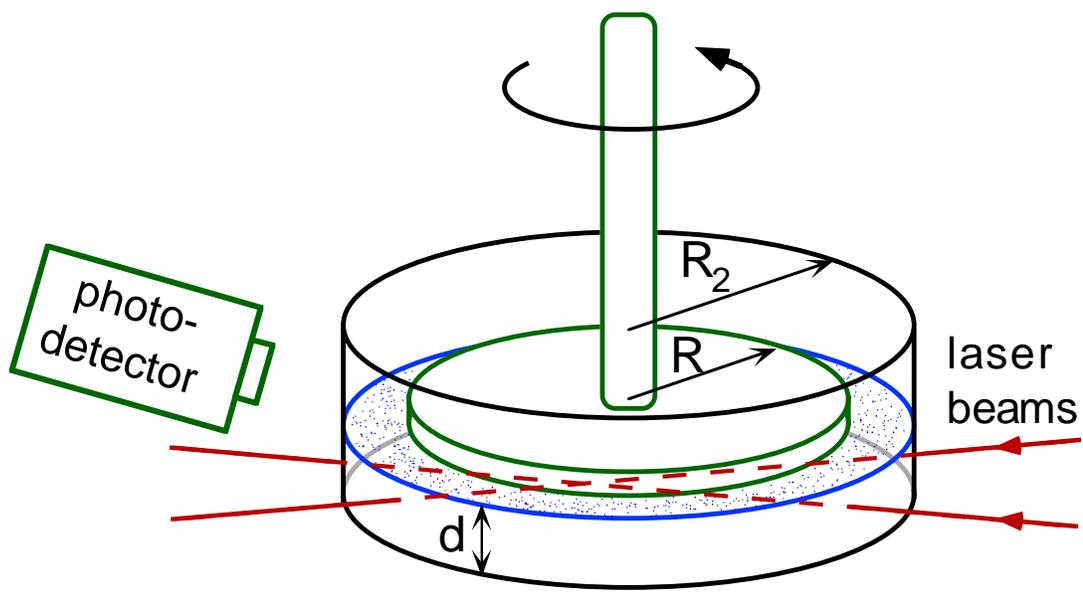
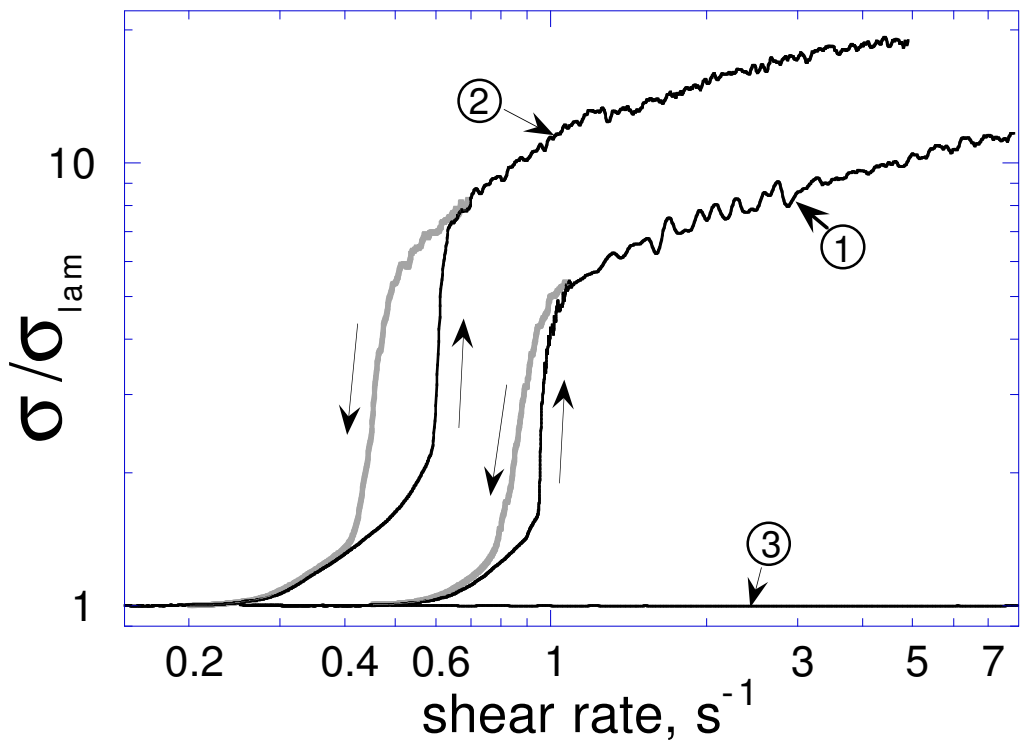
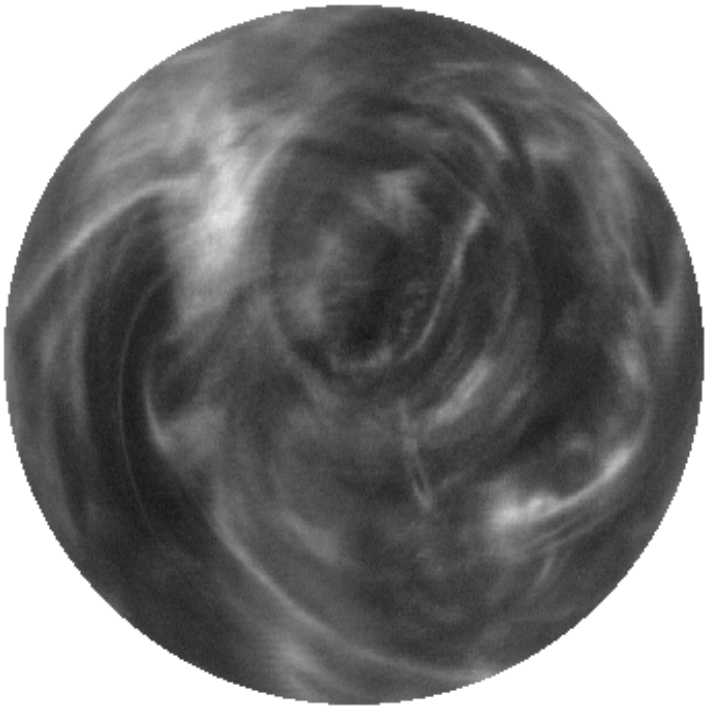
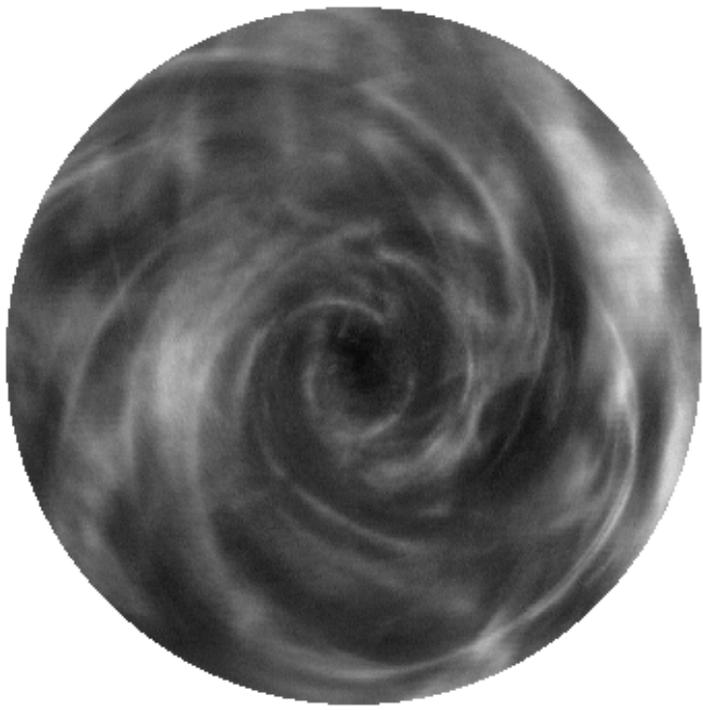
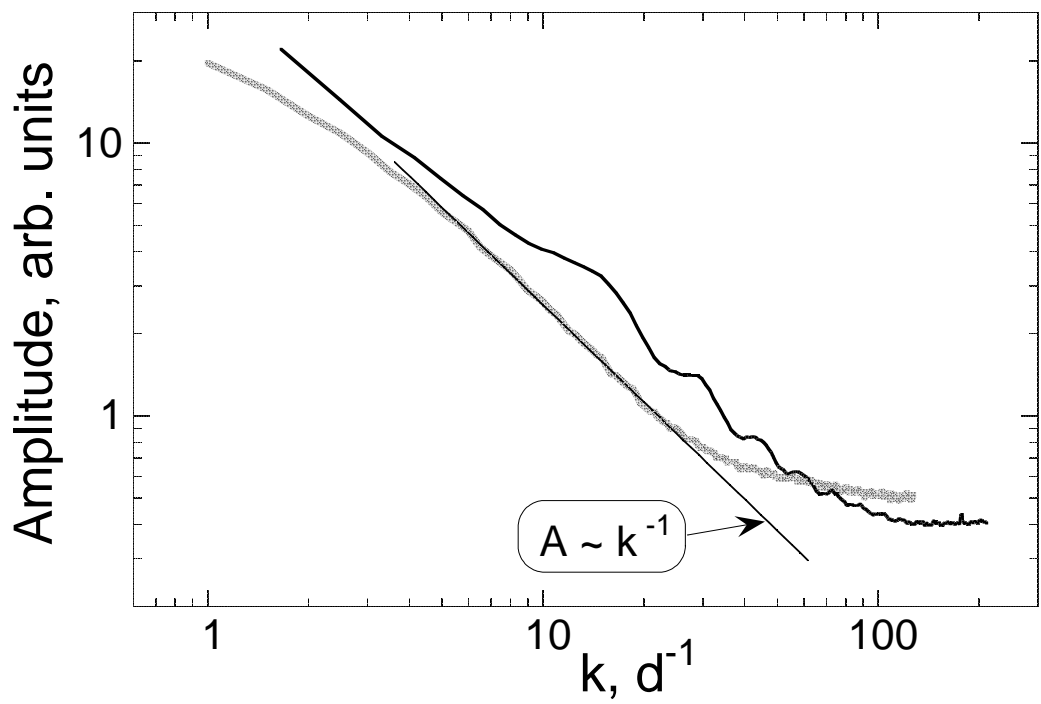
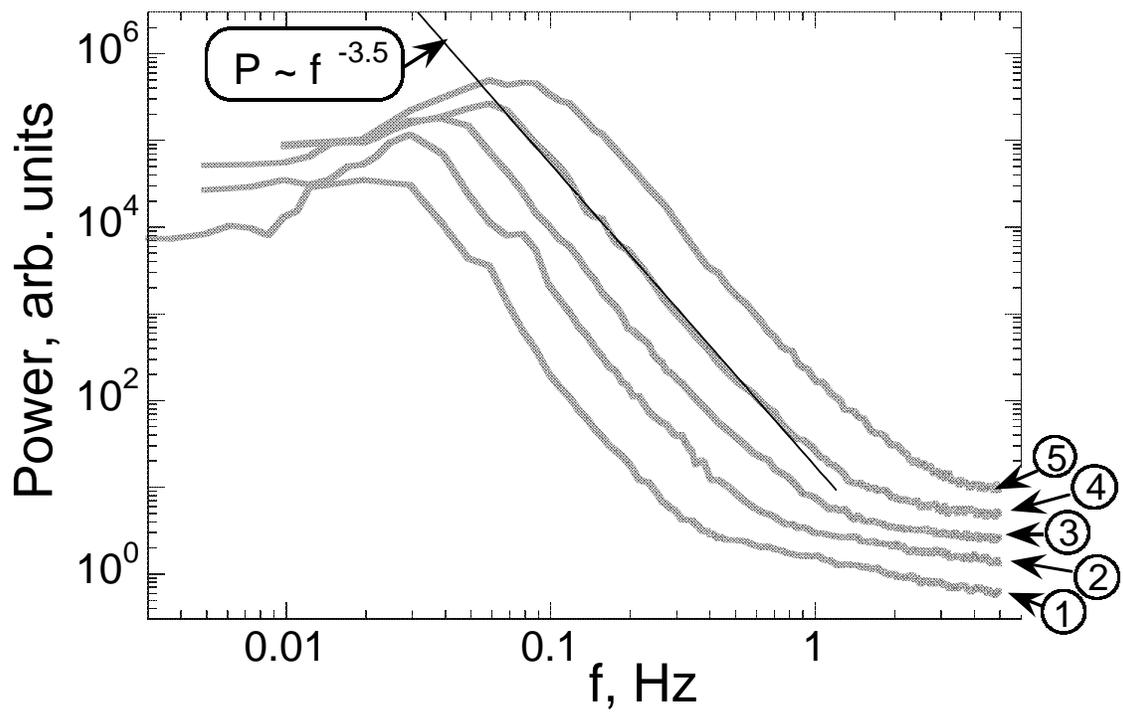
\end{document}